\documentclass[aps,superscriptaddress,amsmath,amssymb,twocolumn,showpacs,floatfix,english,noshowpacs]{revtex4}

\usepackage{url}
\usepackage{bm,bbm}
\usepackage{graphicx}
\usepackage{color}
\usepackage[colorlinks=true, urlcolor=blue, linkcolor=blue, citecolor=blue, pdftex]{hyperref}
\usepackage[english]{babel}

\usepackage{soul}
\usepackage{blindtext}
\usepackage{lipsum}
\usepackage{nicefrac}

\newcommand{\dagga}{{\phantom{\dagger}}}
\begin{document}

\title{Spin-phonon interactions on the kagome lattice: \\ Dirac spin liquid versus valence-bond solids}

\author{Francesco Ferrari}
%\email[]{ferrari@itp.uni-frankfurt.de}
\affiliation{Institute for Theoretical Physics, Goethe University Frankfurt, Max-von-Laue-Stra{\ss}e 1, D-60438 Frankfurt a.M., Germany}
\author{Federico Becca}
\affiliation{Dipartimento di Fisica, Universit\`a di Trieste, Strada Costiera 11, I-34151 Trieste, Italy}
\author{Roser Valent\'\i}
\affiliation{Institute for Theoretical Physics, Goethe University Frankfurt, Max-von-Laue-Stra{\ss}e 1, D-60438 Frankfurt a.M., Germany}

\date{\today}

\begin{abstract}
We investigate the impact of the spin-phonon coupling on the $S=1/2$ Heisenberg model on the kagome lattice. For the pure spin model, there is increasing evidence 
that the low-energy properties can be correctly described by a Dirac spin liquid, in which spinons with a conical dispersion are coupled to emergent gauge fields. 
Within this scenario, the ground-state wave function is well approximated by a Gutzwiller-projected fermionic state [Y. Ran, M. Hermele, P.A. Lee, and X.-G. Wen, 
Phys. Rev. Lett.  {\bf 98}, 117205 (2007)]. However, the existence of $U(1)$ gauge fields may naturally lead to instabilities when small perturbations are included. 
Phonons are ubiquitous in real materials and may play a relevant role in the determination of the actual physical properties of the kagome antiferromagnet. Therefore, 
we perform a step forward in this direction, including phonon degrees of freedom (at the quantum level) and applying a variational approach based upon 
Gutzwiller-projected fermionic {\it Ans\"atze}. Our results suggest that the Dirac spin liquid is {\it stable} for small spin-phonon couplings, while valence-bond 
solids are obtained at large couplings. Even though different distortions can be induced by the spin-phonon interaction, the general aspect is that the energy is 
lowered by maximizing the density of perfect hexagons in the dimerization pattern.
\end{abstract}

\maketitle

\section{Introduction}

One of the longest-standing debates in the field of frustrated magnetism and quantum spin liquids concerns the ground state of the spin-$1/2$ antiferromagnetic Heisenberg 
model on the kagome lattice~\cite{misguich_review}. Drawing from early numerical evidence of the absence of long-range magnetic order~\cite{chalker1992,singh1992,leung1993},
several theoretical predictions have been put forward, based on approximations and numerical methods of different kind. The possibility of a (gapped) quantum spin-liquid 
ground state, derived from a large-$N$ expansion based upon groups with symplectic $Sp(N)$ symmetry was originally proposed in Ref.~\cite{sachdev1991}. On the other hand, 
the (fermionic) $SU(N)$ approximation adopted in Ref.\cite{marston1991} pointed toward the existence of a dimerized ground state, characterized by the largest possible
density of {\it perfect hexagons}, i.e., hexagonal plaquettes with three singlets. The valence-bond ordered picture was further corroborated by the results of other 
numerical methods, e.g., series expansion and quantum dimer models, which identified a dimerized ground state with a supercell of $36$ sites and a honeycomb-like arrangement
of perfect hexagons as energetically most favorable~\cite{nikolic2003,singh2007,poilblanc2011}. This conclusion was strengthened by the calculations of low-lying excitations
above the valence-bond ordered ground state~\cite{singh2008,poilblanc2010}, which reproduced some of the features observed in exact diagonalization calculations on small 
clusters, most notably the large density of singlet excitations below the triplet gap~\cite{lecheminant1997,mila1998,mambrini2000,laeuchli2019}. Interestingly, within a 
short-range valence-bond basis, the lowest-energy singlet excitations were found to be the dimer coverings possessing a large number of perfect hexagons~\cite{zeng1995}. 

More recently, the valence-bond ordered scenario was challenged by various methods, e.g., density-matrix renormalization group (DMRG)~\cite{jiang2008,yan2011,depenbrock2012} 
and variational Monte Carlo~\cite{iqbal2013}, which found a quantum spin liquid ground state in large-scale calculations. Within the latter approach, the best variational 
{\it Ansatz} was found to be a $U(1)$ Dirac state, namely a spin liquid with conical points in the spinon spectrum and emergent $U(1)$ gauge fields in the low-energy 
theory~\cite{ran2007,wen2002}. This state, described by a Gutzwiller-projected fermionic wave function, was shown to be stable to gap-opening instabilities, such as 
dimerization~\cite{iqbal2012} and the lowering of the $U(1)$ gauge structure to $Z_2$~\cite{iqbal2011z2}.

The presence of a Dirac spin liquid ground state is further supported by recent state-of-the-art numerical results. Indeed, although the first DMRG studies indicated the 
presence of a gapped spectrum~\cite{jiang2008,yan2011,depenbrock2012}, more recent calculations on infinitely long cylinders revealed the existence of Dirac spinons by 
means of adiabatic flux insertion~\cite{he2017,zhu2018}, which helped identifying fingerprints of gapless excitations that are gapped by the cylindrical 
geometry~\cite{ferrari2021cylinders}. Furthermore, recent tensor-network calculations also corroborated the gapless nature of the spin liquid ground state~\cite{liao2017}.

From a theoretical perspective, the stability of $U(1)$ Dirac spin liquids is enhanced on non-bipartite geometries, such as the kagome lattice~\cite{song2019}. Still, 
the Dirac spin liquid represents the parent state of certain proximate orders, namely magnetic and valence-bond solid phases, which can result from the condensation of 
monopoles with specific quantum numbers~\cite{hastings2000,hermele2008,song2019,song2020}. These instabilities can be triggered by deviations from the ideal nearest-neighbor
kagome antiferromagnet, e.g., by the inclusion of longer-range exchange interactions~\cite{gong2015,kolley2015,iqbal2021,wietek2020,kiese2023}. In this regard, a 
theoretically and experimentally relevant question concerns the possible spin-Peierls instability of the Dirac spin-liquid state in the presence of a coupling between spins 
and lattice distortions (phonons). Indeed, in one dimension, the gapless spin-liquid ground state of the Heisenberg chain is known to be unstable towards dimerization when 
spins are coupled to {\it static} lattice distortions\cite{cross1979,pytte1974}; however, when phonon degrees of freedom are treated {\it dynamically}, the spin-liquid
phase remains stable below a finite critical value of the spin-phonon coupling~\cite{uhrig1998,bursill1999,sandvik1999,ferrari2020}. In two dimensions the situation is less 
understood and only a smaller number of works have focused on the impact of the spin-phonon coupling on spin liquid phases. For what concerns $Z_2$ spin liquids, there are
studies addressing detectable signatures of spin-lattice coupling in the phonon dynamics of the Kitaev spin liquid~\cite{ye2020,metavitsiadis2022,singh2023}. Furthermore, 
the phonon-driven transition between the $Z_2$ Dirac spin liquid phase on the square lattice and a dimerized phase has been assessed by us in Ref.~\cite{ferrari2021}. 
On the kagome lattice, the spin-Peierls transition has been discussed only in the case of large magnetic fields, for exact eigenstates with localized 
magnons~\cite{richter2004}. However, the question of the stability of the kagome Dirac spin liquid to lattice distortions has not been addressed so far. Its proximity to 
several competing orders within a small energy range~\cite{laeuchli2019} may be interpreted as a sign of potential fragility towards dimerization. In relation to this, 
a recent field theoretical study identified a symmetry-allowed coupling between lattice distortions and monopole operators of the $U(1)$ Dirac spin liquid on the triangular 
lattice as a possible mechanism driving a spin-Peierls transition~\cite{seifert2023}. 

Motivated by these observations, we employ a variational Monte Carlo technique to investigate the ground-state properties of a spin-phonon Hamiltonian on the kagome lattice,
treating the full quantum dynamics of both spins and phonons. We explore several channels of instability of the $U(1)$ Dirac spin liquid towards dimerization, which 
encompass also valence-bond ordered phases associated to its monopole operators~\cite{song2019}. The results of our study indicate that the $U(1)$ Dirac spin liquid is 
stable, i.e., a (first-order) spin-Peierls transition is found to take place only for finite values of the spin-phonon coupling. Then, the valence-bond solids favored by 
the spin-lattice interactions are those which maximize the number of rotationally-symmetric hexagonal plaquettes with strong singlet correlations on their edges.

\section{Model and method}

%%%%%%%%%%%%%%%%%%%%%%%%%%%%%%%%%%%%%%%%%%%%%%%%%%%%%%%%%%%%%%%%%%%%%%%%%%%%%%%%%%%%
\begin{figure}[t]
\includegraphics[width=0.75\columnwidth]{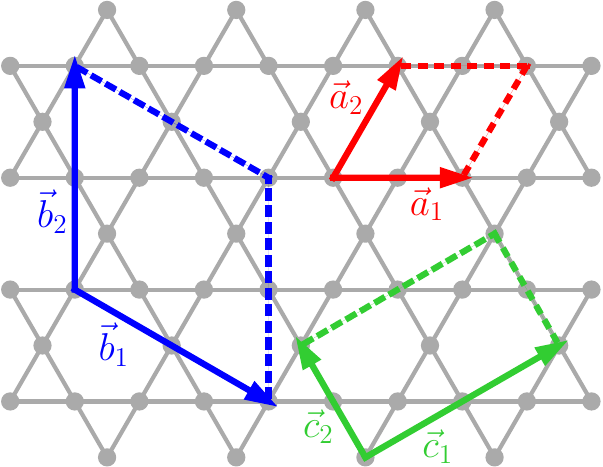}
\caption{\label{fig:supercells}
Lattice vectors employed to construct different supercells for the variational {\it Ans\"atze}. Taking the nearest-neighbor distance as unit, the red vectors are defined 
as ${\vec{a}_1=(2,0)}$ and ${\vec{a}_2=(1,\sqrt{3})}$ (primitive vectors). The blue vectors are ${\vec{b}_1=2\vec{a}_1-\vec{a}_2}$ and ${\vec{b}_2=2\vec{a}_2-\vec{a}_1}$. 
The green vectors are ${\vec{c}_1=\vec{a}_1+\vec{a}_2}$ and ${\vec{c}_2=\vec{a}_2-\vec{a}_1}$.}
\end{figure}
%%%%%%%%%%%%%%%%%%%%%%%%%%%%%%%%%%%%%%%%%%%%%%%%%%%%%%%%%%%%%%%%%%%%%%%%%%%%%%%%%%%%

We consider a system of localized $S=1/2$ spins on the kagome lattice, which are coupled at nearest neighbors by the antiferromagnetic Heisenberg interaction $J$. The 
spin-phonon effects due to lattice distortions are modeled by assuming that the exchange coupling among the spins depends linearly on sites displacements. The system is 
thus described by the following Su-Schrieffer-Heeger~\cite{su1979} Heisenberg Hamiltonian
\begin{align}\label{eq:ham}
&\mathcal{H}=\mathcal{H}_{\mathrm{sp}}+\mathcal{H}_{\mathrm{ph}} \\
&
\mathcal{H}_{\mathrm{sp}}=J\sum_{\langle i,j \rangle}  \left[1-g\frac{\vec{r}_{i}-\vec{r}_j}{ \|\vec{r}_{i}-\vec{r}_{j} \|} \cdot (\vec{U}_i-\vec{U}_j)\right] \mathbf{S}_i \cdot \mathbf{S}_j \\
&\mathcal{H}_{\mathrm{ph}}= \frac{\omega}{4} \sum_j \left(\vec{P}_j^2 + \vec{U}_j^2 \right).
\end{align}
Here $\mathbf{S}_j=(S_j^x,S_j^y,S_j^z)$ denotes the spin operator at site $j$. The lattice degrees of freedom are described by a set of local harmonic oscillators through 
the displacement and momentum operators, ${\vec{U}_j=(X_j,Y_j)}$ and ${\vec{P}_j=(P^X_j,P^Y_j)=-2i(\partial_{X_j},\partial_{Y_j})}$. Within this 
notation~\cite{ferrari2020,ferrari2021}, the canonical commutation relations between phonon operators become $[X_j,P^X_l]=[Y_j,P^Y_l]=2i\delta_{j,l}$ (all other commutators 
vanish). The strength of the spin-phonon coupling is controlled by the adimensional constant $g$. The vector $\vec{r}_{i}=(x_i,y_i)$ marks the position of site $i$ in the 
undistorted kagome lattice, and $\|\cdot \|$ denotes the Euclidean norm. The form of the spin-phonon coupling in $\mathcal{H}_{\rm sp}$ arises from the linear term in the 
Taylor expansion of the exchange interaction as a function of the Euclidean distance between sites. The second term of Eq.~\eqref{eq:ham}, $\mathcal{H}_{\mathrm{ph}}$, is 
the free Hamiltonian of Einstein (optical) phonons, i.e., uncoupled harmonic oscillators with a flat dispersion with frequency $\omega$. 

We notice that the model of Eq.~\eqref{eq:ham} is invariant under the transformation $g\mapsto -g$ and ${\vec U}_j \mapsto -{\vec U}_j$ for all the sites $j$; thus, the 
sign of $g$ does not affect the ground-state energy and the phase diagram. Here, we take ${g>0}$, which corresponds to the more physical situation in which negative 
spin-spin correlations (e.g., singlet states) tend to shrink bonds, or, in other words, in which the exchange coupling decreases when spins are pulled far apart.

\subsection*{Variational wave function}

We make use of a variational Monte Carlo method to approximate the ground-state wave function of the spin-phonon problem~\eqref{eq:ham} on finite-size clusters with $N$ 
sites and periodic boundary conditions. Our variational {\it Ansatz} is the product of a spin state $|\Psi_s\rangle$, a phonon state $|\Psi_p\rangle$, and a Jastrow 
factor $\mathcal{J}_{sp}$ which correlates spins with lattice distortions~\cite{ferrari2020,ferrari2021,ferrari2022}:
\begin{equation}\label{eq:psi0}
|\Psi_{\mathrm{var}}\rangle=\mathcal{J}_{sp} |\Psi_p\rangle \otimes |\Psi_s\rangle.
\end{equation}
The variational energy $E_{\rm var}$ is estimated by performing a Markov process in the total Hilbert space that includes spin and phonon configurations 
$|\vec{u}; s^z \rangle = \bigotimes_j (|\vec{u}_j\rangle \otimes  |s^z_j \rangle)$, i.e., in the local eigenbasis of the ${\vec U}_j$ (phonon) and $S_j^z$ (spin) operators 
for each lattice site $j$~\cite{ferrari2020}. Then, we have:

\begin{align}\label{eq:varener}
E_{\mathrm{var}} &= \frac{\langle \Psi_{\mathrm{var}}| \mathcal{H} |\Psi_{\mathrm{var}}\rangle}{\langle \Psi_{\mathrm{var}}|\Psi_{\mathrm{var}}\rangle} \nonumber \\ &= 
\sum_{s^z} \int d \vec{u} \, \underbrace{\frac{|\langle \vec{u}; s^z|\Psi_{\mathrm{var}}\rangle|^2}{\langle \Psi_{\mathrm{var}}|\Psi_{\mathrm{var}}\rangle}}_{{\cal P}(\vec{u}; s^z)} \,
\underbrace{\frac{\langle \vec{u}; s^z| \mathcal{H} |\Psi_{\mathrm{var}}\rangle}{\langle \vec{u}; s^z|\Psi_{\mathrm{var}}\rangle}}_{e_L(\vec{u}; s^z)}
\end{align}
where the sum is over all spin configurations and the integral is over all phonon displacements. Then, along the Markov chain, a set of configurations $(\vec{u}; s^z)_m$ 
(with $m=1,\dots,M$) are drawn according to the probability ${\cal P}(\vec{u}; s^z)$ and the Metropolis algorithm, which allows us to estimate the variational energy as:
\begin{equation}
E_{\mathrm{var}} \approx \frac{1}{M} \sum_{m=1}^M e_L(\vec{u}; s^z)_m.
\end{equation}
The stochastic process is performed in the subspace of zero total spin along $z$ and the Metropolis moves for the spins involve  nearest-neighbor double spin-flips; for the phonon degrees of freedom, the Metropolis moves consist of local updates of the displacements, $X_j \mapsto X_j +\Delta$ or $Y_j \mapsto Y_j +\Delta$, with $\Delta$ uniformly distributed within a certain interval $[-\Delta_{\rm max},\Delta_{\rm max}]$.

The form of the variational state is partially dictated by the efficiency in the calculation of its amplitudes $\langle \vec{u}; s^z|\Psi_{\mathrm{var}}\rangle$. In this
respect, the spin-phonon Jastrow factor must be diagonal in the chosen computational basis (otherwise it would require the calculations of matrix elements with all states of the basis). 
Therefore, we are limited to consider terms involving the $z$-component of the spin operators, thus breaking the SU(2) symmetry, and take~\cite{ferrari2020}:
\begin{equation}
\mathcal{J}_{sp}=\exp\left[\frac{1}{2}\sum_{i,j}  \vec{v}_{sp}(i,j)\cdot(\vec{U}_i-\vec{U}_j) S^z_i S^z_j\right],
\end{equation}
where the pseudopotential parameters (antisymmetric under the exchange of $i$ and $j$ indices) are defined as
\begin{equation}
  \vec{v}_{sp}(i,j)=\begin{pmatrix}
v_{sp}^X(\|\vec{r}_{i}-\vec{r}_{j} \|) \frac{x_i-x_j}{\|\vec{r}_{i}-\vec{r}_{j} \|} \\[0.25cm]
v_{sp}^Y(\|\vec{r}_{i}-\vec{r}_{j} \|) \frac{y_i-y_j}{\|\vec{r}_{i}-\vec{r}_{j} \|}
\end{pmatrix}.
\end{equation}

Then, given the {\it Ansatz} of Eq.~\eqref{eq:psi0}, we have that
\begin{equation}
\langle \vec{u}; s^z|\Psi_{\mathrm{var}}\rangle = J_{sp}(\vec{u}; s^z) \, \langle \vec{u}|\Psi_p\rangle \, \langle s^z|\Psi_s\rangle,
\end{equation}
where $J_{sp}(\vec{u}; s^z)$ is the value that the operator $\mathcal{J}_{sp}$ acquires on the configuration $|\vec{u}; s^z \rangle$. The phonon part of the variational 
{\it Ansatz} is a coherent state:
\begin{equation}\label{eq:psip}
  \langle \vec{u} |\Psi_p\rangle= \prod_i \exp\left(-\frac{\|\vec{u}_i-\vec{z}_i \|^2}{4}\right),
\end{equation}
which is written as a product of Gaussians that are centered in the displaced positions of the lattice sites, represented by the variational parameters 
$\vec{z}_i=(z^X_i,z^Y_i)$. In general, non-zero values of $\vec{z}_i$ induce the presence of finite lattice distortion. Finally, the spin part of the variational state 
is defined within the Abrikosov fermion representation of $S=1/2$ spin operators~\cite{abrikosov1965,savary2016}
\begin{equation}
\mathbf{S}_i=\frac{1}{2} \sum_{\sigma,\sigma^\prime} c^\dagger_{i,\sigma} \boldsymbol{\sigma}_{\sigma,\sigma^\prime} c^\dagga_{i,\sigma^\prime},
\end{equation}
where $c^\dagga_{i,\sigma}$ ($c^\dagger_{i,\sigma}$) is the annihilation (creation) operator of a fermion at site $i$ with spin $\sigma=\uparrow, \downarrow$, and 
$\boldsymbol{\sigma}=(\sigma_x,\sigma_y,\sigma_z)$ is a vector of Pauli matrices. The state $|\Psi_s\rangle$ consists of a fermionic Slater determinant, 
$|\Phi_0\rangle$, which is projected onto spin space by the Gutzwiller operator ${\mathcal{P}_G=\prod_i (2-n_i)n_i}$, where 
$n_i=\sum_\sigma c^\dagger_{i,\sigma}c^\dagga_{i,\sigma}$, and supplemented by a spin Jastrow factor $\mathcal{J}_{ss}$ (diagonal in the computational
basis):
\begin{equation}
|\Psi_s\rangle=\mathcal{J}_{ss}\mathcal{P}_G|\Phi_0\rangle.
\end{equation}
We define $|\Phi_0\rangle$ as the ground state of an auxiliary tight-binding Hamiltonian on the kagome lattice, with nearest-neighbor hopping terms  
\begin{equation}\label{eq:hamH0}
{\mathcal{H}}_{\rm 0} = \sum_{\langle i,j \rangle} \sum_\sigma t_{i,j} c^\dagger_{i,\sigma} c^\dagga_{j,\sigma} + {\rm h.c.},
\end{equation}
and the Jastrow factor as
\begin{equation}
\mathcal{J}_{ss}=\exp\left[\frac{1}{2}\sum_{i,j} v_{ss}(\|\vec{r}_i-\vec{r}_j\|) S_i^z S_j^z\right].
\end{equation}
The Gutzwiller projection is implemented by the Markov chain, in which spin configurations are sampled, i.e., fermionic configurations with singly occupied sites.
Then, we have that
\begin{equation}
\langle s^z|\Psi_s\rangle = J_{ss}(s^z) \, {\mathrm{det}} (\phi_{r_{\uparrow},\alpha}) \, {\mathrm{det}} (\phi_{r_{\downarrow},\alpha})
\end{equation}
where $\phi_{r_{\sigma},\alpha}$ are the $N/2$ (spin-independent) lowest-energy orbitals of the auxiliary Hamiltonian~\eqref{eq:hamH0}, labelled by the index $\alpha$ 
and evaluated on the positions of up ($r_\uparrow$) and down spins ($r_\downarrow$) in the configuration $|s^z \rangle$; $J_{ss}(s^z)$ is the value of the operator 
$\mathcal{J}_{ss}$ on the configuration $|s^z\rangle$. 

The parameters, which are optimized in order to minimize the variational energy~\eqref{eq:varener}, are given by the hoppings ($t_{i,j}$) of the auxiliary Hamiltonian, 
the centers of the Gaussians ($\vec{z}_i$), and the Jastrow pseudopotentials [$v_{ss}(\|\vec{r}_i-\vec{r}_j\|)$, $v_{sp}^X(\|\vec{r}_{i}-\vec{r}_{j} \|)$, and 
$v_{sp}^Y(\|\vec{r}_{i}-\vec{r}_{j} \|)$] at all inequivalent lattice distances. The optimization procedure is performed by using the stochastic reconfiguration 
method~\cite{sorella2005}. 

%%%%%%%%%%%%%%%%%%%%%%%%%%%%%%%%%%%%%%%%%%%%%%%%%%%%%%%%%%%%%%%%%%%%%%%%%%%%%%%%%%%%
\begin{figure*}[t]
\includegraphics[width=\textwidth]{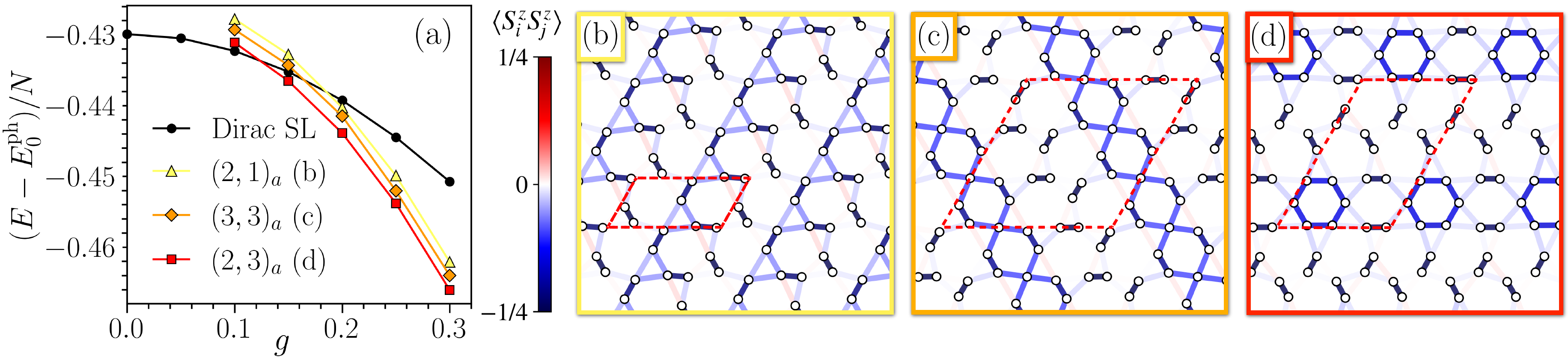}
\caption{\label{fig:6x6ene}
Results of the calculations on the $(6,6)_a$ cluster ($N=108$ sites). In panel (a) the energies (per site) of the Dirac spin liquid (SL) and different valence-bond 
ordered states are shown as a function of the spin-phonon coupling $g$. The zero-point energy of the phonon Hamiltonian, $E_0^{\rm ph}=N\omega$, has been subtracted. 
The patterns of distortions associated to the various valence-bond ordered states are shown in panels (b),(c),(d), as obtained within the $(2,1)_a$, $(3,3)_a$, $(2,3)_a$ 
supercells, respectively. The color of the bonds between sites $i$ and $j$ represents the value of the spin-spin correlations $\langle S^z_i S^z_j\rangle$ (at $g=0.3$). 
The dashed lines delimit the supercells.}
\end{figure*}
%%%%%%%%%%%%%%%%%%%%%%%%%%%%%%%%%%%%%%%%%%%%%%%%%%%%%%%%%%%%%%%%%%%%%%%%%%%%%%%%%%%%

\subsection*{Lattice distortions and supercells}

A challenging aspect of this study is determining the most favorable lattice distortions that are induced by the spin-phonon interaction and could represent potential 
instabilities of the Dirac spin liquid state. Instead of considering a variational phonon {\it Ansatz} $|\Psi_p\rangle$ with a fixed phonon momentum and a certain 
polarization (as done in Ref.~\cite{ferrari2021}), we apply a more general approach in which we scan through several supercells that can accommodate different patterns 
of distortions. Within this method, the variational parameters $t_{i,j}$ [Eq.~\eqref{eq:hamH0}] and $\vec{z}_i$ [Eq.~\eqref{eq:psip}] are taken to be translationally 
invariant with respect to the chosen supercell. Thus, for a supercell of $N_s$ sites, $2 N_s$ independent nearest-neighbor hoppings and $N_s$ $\vec{z}_i$ parameters are 
optimized. This strategy allows us to describe combinations of phonon modes with different momenta and polarizations. 

We consider supercells of various sizes and shapes. They are defined by means of three distinct sets of vectors, $\vec{a}_{i}$, $\vec{b}_i$ and $\vec{c}_i$ ($i=1,2$), 
shown in Fig.~\ref{fig:supercells}. The supercells constructed by the $a$-vectors are denoted as $(m,n)_a$ ($m,n \in\mathbb{N}$) and defined by the vectors $m \vec{a}_1$ 
and $n \vec{a}_2$. An analogous notation is used for the supercells constructed by the $\vec{b}_i$ and $\vec{c}_i$ vectors. We note that the $(m,n)_{a,b,c}$ supercells 
contain $3 m \times n$, $9 m \times n$ and $6 m \times n$ sites, respectively. We scan several supercells described by these vectors, up to a size of $48$ sites. The 
full cluster, on which numerical calculations are performed, will be also denoted by a similar notation. The three families of supercells are chosen because they 
represent simple geometries that can be fitted inside the finite-size clusters which are accessible to numerical simulations. In addition, the supercells defined by 
$\vec{a}_{i}$ or $\vec{b}_i$ vectors fulfill all the point-group symmetries of the original kagome lattice if $m=n$, while the one defined by $\vec{c}_i$ vectors has 
been discussed in Ref.~\cite{nikolic2003}.

For what concerns the fermionic variational state, we initialize the hoppings of the auxiliary Hamiltonian $\mathcal{H}_0$ in the vicinity of one of the four $U(1)$ 
symmetric spin liquids of the kagome lattice~\cite{lu2011}. Specifically, the initial signs of the hopping parameters reproduce the $0$ and/or $\pi$ fluxes through the 
triangular and hexagonal plaquettes which characterize the different $U(1)$ spin liquids~\cite{lu2011}. The idea behind this approach is motivated by the fact that a 
symmetric spin liquid can serve as a parent state of certain valence-bond solid instabilities~\cite{hastings2000,hermele2008,iqbal2012,song2019,ferrari2021,kiese2023}.
However, since all hopping parameters within the supercell are independently optimized, they can take any value upon energy minimization and thus change the fluxes from 
their initial values.

\section{Results}

We study the model of Eq.~\eqref{eq:ham} for $\omega/J=1$ and different values of the spin-phonon coupling $g$. Throughout the paper, we set $J=1$ to fix the energy scale 
of the problem. At $g=0$ the lattice is undistorted and the optimal variational state is given by the $U(1)$ Dirac spin liquid state~\cite{ran2007,iqbal2013}. Turning on 
the spin-phonon interaction, we observe the appearance of several competing valence-bond ordered states, whose energies eventually beat the one of the Dirac spin liquid 
for large values of $g$. The main findings of this paper can be summarized in three points: i) the Dirac spin liquid state is {\it stable} for small values of the 
spin-phonon interaction; ii) the valence-bond ordered states which provide the lowest energy for large $g$ are characterized by the appearance of hexagonal plaquettes 
with strong spin-spin correlations; iii) the transition between the Dirac spin liquid and the best distorted states is of the first order. 

We begin the discussion by looking at the results obtained on the $(6,6)_a$ cluster (with $N=108$ sites), which exemplify the main observations outlined above. For the sake 
of clarity, we restrict our analysis to $(m,n)_a$ supercells. As shown in Fig.~\ref{fig:6x6ene}(a), for $g\lesssim 0.1$, the Dirac spin liquid gives the best variational 
energy, but several metastable valence-bond solids with finite distortions can be stabilizied within different supercells. When the spin-phonon coupling is sufficiently 
strong, all these dimerized states become lower in energy than the Dirac spin liquid. The different distortions are shown in Fig.~\ref{fig:6x6ene}(b-d), where the color 
of the bonds between neighboring sites represents the value of the spin-spin correlations. The higher-energy distorted state [$(2,1)_a$ in Fig.~\ref{fig:6x6ene}(b)] 
is essentially made of dimers which form stripes along $a_2$. The lowest-energy state, instead, is obtained by the $(2,3)_a$ supercell [Fig.~\ref{fig:6x6ene}(d)] and, in 
addition to stripes of alternating dimers, shows the appearance of hexagonal plaquettes with strong antiferromagnetic sides. We dub these rotationally-symmetric plaquettes 
{\it perfect hexagons}. This terminology is borrowed from studies concerning (hard-core) dimer coverings on the kagome lattice, where perfect hexagons are hexagonal 
plaquettes, which host three dimers and are surrounded by empty triangular plaquettes~\cite{marston1991,zeng1995,nikolic2003,singh2007,poilblanc2010}. In the case of 
hard-core dimers, due to geometrical constraints, the maximum density of perfect hexagons among hexagonal plaquettes is found to be $1/6$, i.e., one every $18$ sites. 
The $(2,3)_a$ valence-bond solid in Fig.~\ref{fig:6x6ene} is analogous to the $18$-sites dimer covering found by Marston and Zeng within a large-$N$ expansion of the 
kagome antiferromagnet, which shows the maximum density of perfect hexagons~\cite{marston1991}. The difference lies in the fact that in the present case, in absence of the 
hard-core dimer constraint, perfect hexagons possess six edges with equal spin-spin correlations, instead of having three strong and three weak bonds. It is worth 
mentioning that an intermediate energy state is found by considering the $(3,3)_a$ supercell. As shown in Fig.~\ref{fig:6x6ene}(c), this valence-bond solid is characterized 
by parallel stripes of dimers which are connected by relatively strong antiferromagnetic correlations, forming lines of hexagons. We note, however, that these hexagons are 
not perfect, as their edges do not possess equally strong spin-spin correlations. 

%%%%%%%%%%%%%%%%%%%%%%%%%%%%%%%%%%%%%%%%%%%%%%%%%%%%%%%%%%%%%%%%%%%%%%%%%%%%%%%%%%%%
\begin{figure*}[t]
\includegraphics[width=0.9\textwidth]{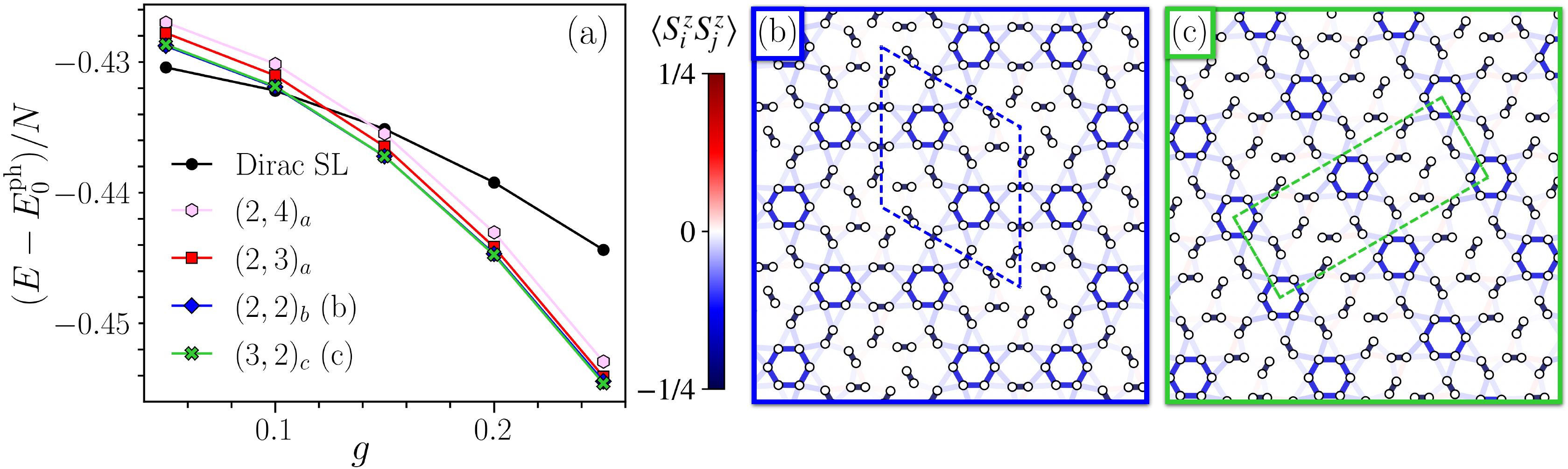}
\caption{\label{fig:12x12ene}
The same as in Fig.~\ref{fig:6x6ene} for calculations on the $(12,12)_a$ cluster ($N=432$ sites). Panel (a) shows the energies (per site) of the Dirac spin liquid (SL) 
and different valence-bond solids with perfect hexagons. Panels (b) and (c) show the distortions and spin-spin correlations (at $g=0.3$) of the $(2,2)_b$ and $(3,2)_c$ 
supercells, respectively.}
\end{figure*}
%%%%%%%%%%%%%%%%%%%%%%%%%%%%%%%%%%%%%%%%%%%%%%%%%%%%%%%%%%%%%%%%%%%%%%%%%%%%%%%%%%%%

The preliminary results on the $(6,6)_a$ cluster already suggest that a large density of perfect hexagons turns out to be the distinguishing hallmark of the best dimerized 
states of the spin-phonon model~\eqref{eq:ham}. For the complete search of possible patterns of distortions, we carried out several calculations on finite-size clusters of 
various shapes that can accommodate different supercells. The lowest-energy distortions found by this analysis display perfect hexagons in their spin-spin correlation 
patterns. The energies of these states can be compared by performing calculations on the $(12,12)_a$ cluster (with $N=432$ sites), which can accommodate all the optimal 
supercells we identified. The results are summarized in Fig.~\ref{fig:12x12ene}, where different patterns with perfect hexagons are compared. Similarly to what has been 
previously discussed, we find the Dirac spin liquid to be stable to dimerization until a certain critical value of the spin-phonon coupling is reached ($g_c\approx 0.1$). 
For larger values of $g$, the lowest-energy states are given by the $(2,2)_b$ and $(3,2)_c$ supercells, both containing $36$ sites. The lattice distortions and the relative 
spin-spin correlations of these states are shown in Fig.~\ref{fig:12x12ene}(b,c) and display the maximal density of perfect hexagons (two per supercell, i.e., one every 
$18$ sites). Both these patterns have been discussed in several works as possible ground states of spin (or quantum dimer) models on the kagome 
lattice~\cite{marston1991,zeng1995,nikolic2003,singh2007,poilblanc2010,evenbly2010,poilblanc2011,huh2011,iqbal2012}. The valence-bond order of Fig.~\ref{fig:12x12ene}(b) 
shows a honeycomb structure of perfect hexagons, which are arranged around weak hexagonal plaquettes surrounded by a {\it pinwheel} pattern of dimers. Previous studies 
on kagome Heisenberg models showed that this valence-bond solid can be favored by a ferromagnetic second-neighbor coupling~\cite{iqbal2012}. The distortion in 
Fig.~\ref{fig:12x12ene}(c), instead, is characterized by parallel stripes of perfect hexagons~\cite{nikolic2003}. On the $(12,12)_a$ cluster, the energy obtained by the 
two $36$-sites supercells, $(2,2)_b$ and $(3,2)_c$, are slightly lower than the one of the $18$-sites supercell $(2,3)_a$ of Fig.~\ref{fig:6x6ene}(d), at least for small 
values of the spin-phonon coupling. It is not possible, instead, to draw definitive conclusions about which of the two $36$-sites valence-bond orders is better than the 
other, as the relative energy difference is very small and difficult to resolve within stochastic and optimization uncertainties. Still, it can be clearly stated that a 
large density of perfect hexagons is crucial to obtain the lowest energies. Indeed, similarly to the correlation pattern in Fig.~\ref{fig:6x6ene}(d), the valence-bond 
state obtained with the $(2,4)_a$ supercell is also characterized by perfect hexagons separated by stripes of alternating dimers (not shown). However, it possesses a lower 
density of perfect hexagons, i.e., one every $24$ sites, and its energy is found to be consistently higher than the states with maximal density, as shown in 
Fig.~\ref{fig:12x12ene}(a). 

%%%%%%%%%%%%%%%%%%%%%%%%%%%%%%%%%%%%%%%%%%%%%%%%%%%%%%%%%%%%%%%%%%%%%%%%%%%%%%%%%%%%
\begin{figure}[h]
\includegraphics[width=0.65\columnwidth]{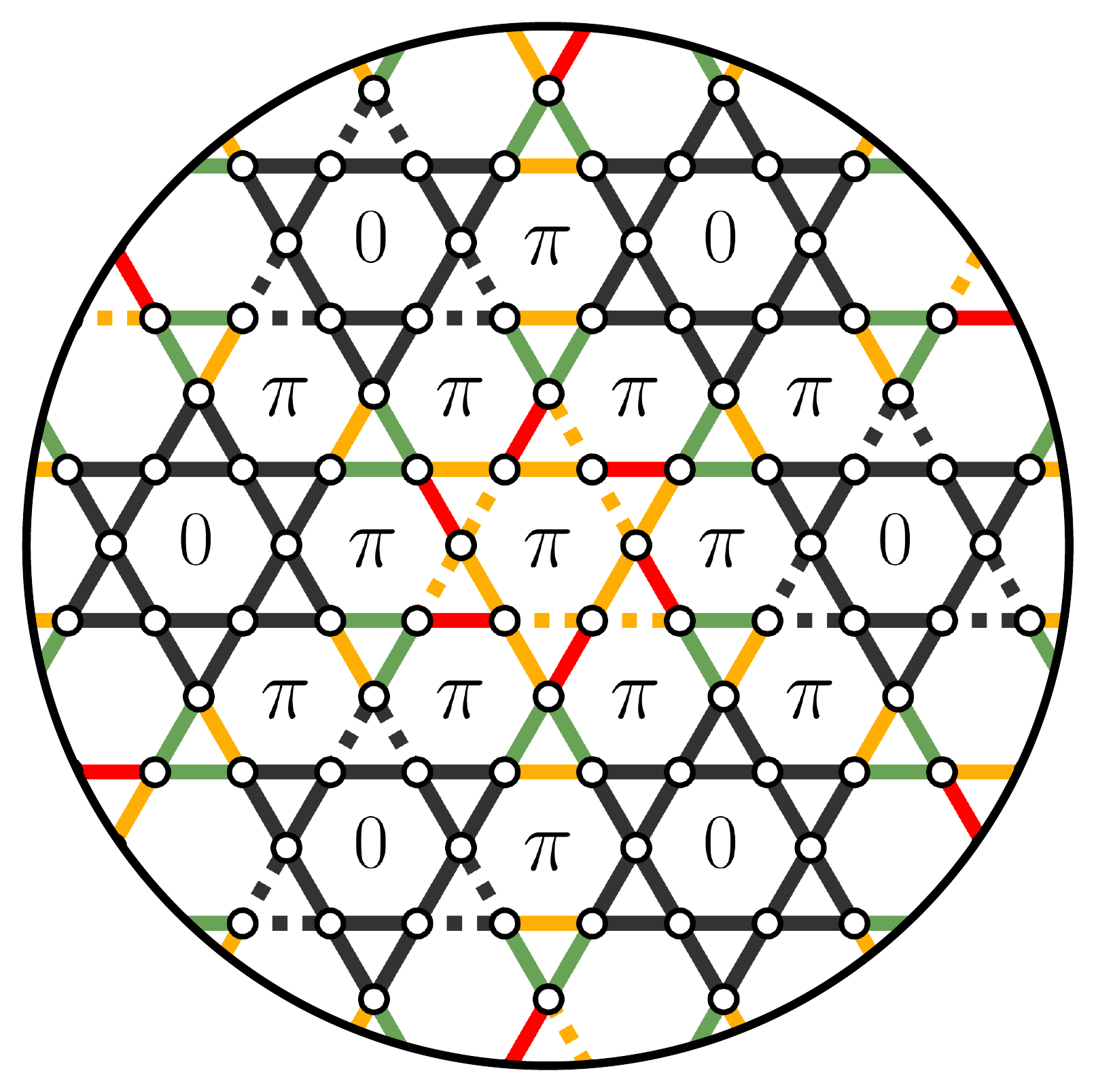}
\caption{\label{fig:fluxes}
Fermionic hoppings in the minimal variational {\it Ansatz} for the $(2,2)_b$ valence-bond solid. Bonds of different colors correspond to (four) independent absolute values 
of the hopping parameters. Solid and dashed lines denote positive and negative signs of the hopping, respectively. The fluxes threading the hexagonal plaquettes are 
reported and compatible with those observed in Ref.~\cite{iqbal2012}. Perfect hexagons can be recognized as the ones with $0$ flux.}
\end{figure}
%%%%%%%%%%%%%%%%%%%%%%%%%%%%%%%%%%%%%%%%%%%%%%%%%%%%%%%%%%%%%%%%%%%%%%%%%%%%%%%%%%%%

The transition between the Dirac spin liquid and the valence-bond-solid states appears to be of the first order. This conclusion is motivated by two different observations. 
On the one hand, the dimerized states can be stabilized as metastable states in the region where the Dirac spin liquid gives the lowest-energy variational {\it Ansatz}. 
On the other hand, the analysis of the fluxes threading the elementary plaquettes of the lattice suggests that the valence-bond solids with perfect hexagons cannot be 
smoothly connected to the Dirac spin-liquid state. Indeed, we find that the product of hoppings around a perfect hexagon always produces $0$-flux, which is incompatible 
with the $\pi$-flux threading the hexagonal plaquettes in the Dirac spin liquid {\it Ansatz}. We also observe that the hexagons which lie right between two perfect hexagons 
(and possess two dimers at opposite edges) are always threaded by flux $\pi$. No general conclusions can be drawn for the fluxes of the other hexagonal plaquettes of the 
lattice.

\begin{figure}
\includegraphics[width=\columnwidth]{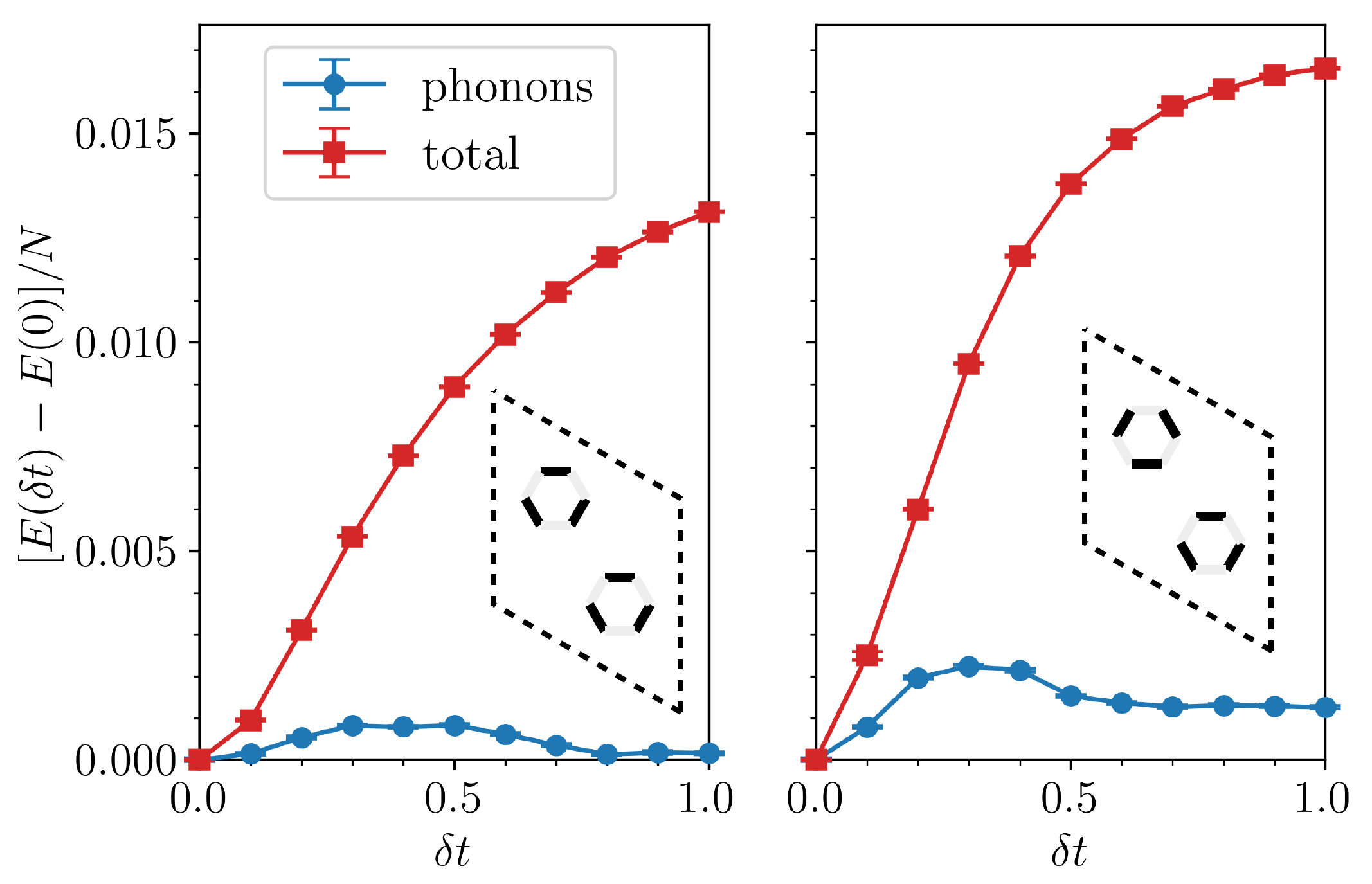}
\caption{\label{fig:which_ene}
Energy loss due to forcing a dimerization of the perfect hexagons in the variational {\it Ansatz} of the $(2,2)_b$ valence-bond solid (Fig.~\ref{fig:fluxes}). The total 
energy loss (red squares) is plotted alongside the contribution due to the free phonon Hamiltonian (blue cirlces), for two inequivalent dimerization patterns (illustrated 
in the insets). The results are obtained on the $(4,4)_b$ cluster ($N=144$ sites), for $g=0.3$.}
\end{figure}

Starting from the results of the unbiased optimization of all the hopping parameters, we found a simpler parametrization for the {\it Ansatz} of the $36$-sites valence-bond 
solid defined by the $(2,2)_b$ supercell [Fig.~\ref{fig:12x12ene}~(b)]. Indeed, the number of independent hoppings can be narrowed down from $72$ to $4$, without worsening 
the variational energy. However, as previously mentioned, the choice of the signs of the hoppings on the different bonds is crucial and induces a specific flux pattern, shown 
in Fig.~\ref{fig:fluxes}. We make use of this simplified variational {\it Ansatz} to perform a numerical experiment to understand why the system prefers to form distortions 
with perfect hexagons with rotationally symmetric pattern of correlations, instead of creating strong dimers on three edges. For this purpose, we start from the variational 
{\it Ansatz} in Fig.~\ref{fig:fluxes}, and we force the hoppings on the even/odd edges of the perfect hexagons to take values $1+\delta t$ and $1-\delta t$, respectively. 
The remaining hoppings are optimized, together with the other variational parameters. As a result, we are able to draw an energy landscape as we smoothly tune the symmetric 
hexagons ($\delta t=0$) into plaquettes with three dimers ($\delta=1$). Two inequivalent dimerizations can be considered, since the supercell contains two perfect hexagons. 
The resulting energy loss is reported in Fig.~\ref{fig:which_ene}, where both the total energy and the contribution from the free phonon Hamiltonian $\mathcal{H}_{\rm ph}$ 
are shown. The latter is found to be a small fraction of the total energy loss. Therefore, we conclude that the tendency to form symmetric perfect hexagons (instead of 
disconnected dimers) is driven by the spin-phonon part of the Hamiltonian.

%%%%%%%%%%%%%%%%%%%%%%%%%%%%%%%%%%%%%%%%%%%%%%%%%%%%%%%%%%%%%%%%%%%%%%%%%%%%%%%%%%%%

\begin{figure}
\includegraphics[width=\columnwidth]{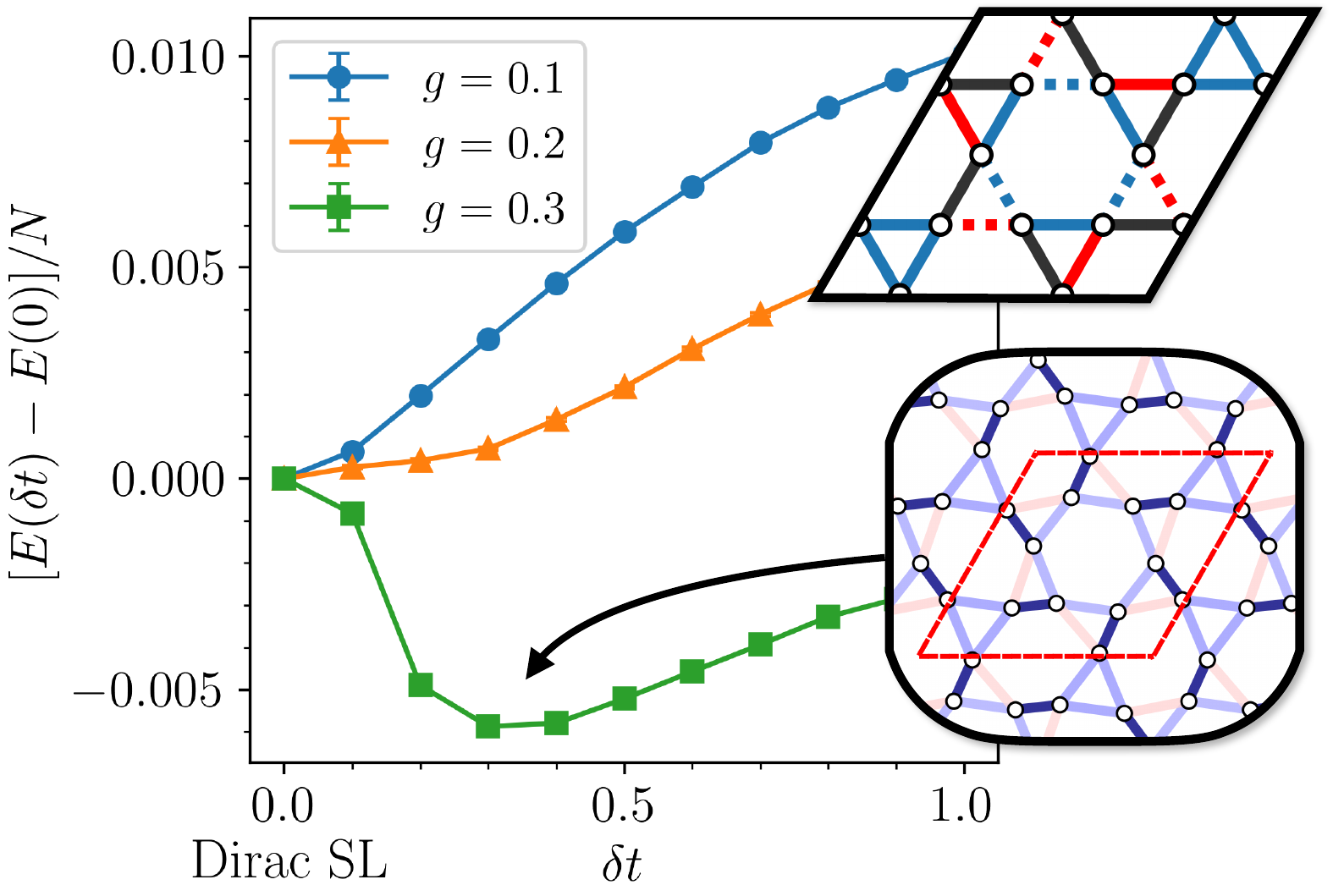}
\caption{\label{fig:pw}
Energy landscape of the {\it pinwheel} valence-bond solid {\it Ansatz}~\cite{song2019} as a function of the tuning parameter $\delta t$. The three independent hoppings 
defining the variational state are shown in the upper inset. Different colors refer to different absolute values, with red/black bonds indicating $|t_{i,j}|=1\pm\delta t$. 
Solid and dashed lines denote, respectively, positive and negative hoppings, which reproduce the characteristic fluxes of the Dirac spin liquid. The lower inset shows the 
lattice distortions and spin-spin correlations (analogously to Fig.~\ref{fig:6x6ene} and~\ref{fig:12x12ene}) in the minimum of the variational energy, for $g=0.3$. 
The results are obtained on the $(6,6)_a$ cluster ($N=108$ sites).}
\end{figure}
%%%%%%%%%%%%%%%%%%%%%%%%%%%%%%%%%%%%%%%%%%%%%%%%%%%%%%%%%%%%%%%%%%%%%%%%%%%%%%%%%%%%

Finally, we mention that finding valence-bond solid instabilities that are continuously connected to the Dirac spin liquid is possible, although they yield higher 
variational energies than those of the optimal distortions presented above. For example, we constructed a variational {\it Ansatz} within the $(2,2)_a$ supercell that 
reproduces the {\it pinwheel} valence-bond-solid pattern, which can arise from monopole condensation and lead to a finite gap in the Dirac spectrum~\cite{song2019}. 
The fermionic Hamiltonian of the variational state contains three hoppings, whose signs yield the characteristic fluxes of the Dirac state, as shown in the upper inset of 
Fig.~\ref{fig:pw}. The absolute values of the hoppings in red and black are fixed to $1+\delta t$ and $1-\delta t$, respectively, while the remaining hopping is optimized, 
together with all the other variational parameters. In this way, we can draw an energy landscape as a function of the $\delta t$ parameter, which induces a pinwheel dimerization 
around the hexagon in the middle of the supercell. As shown in Fig.~\ref{fig:pw}, for small values of the spin-phonon couplings (i.e., $g=0.1$ and $0.2$) the minimum of 
the energy is found at $\delta t=0$. Here, the absolute value of the third hopping converges to $1$, yielding the Dirac spin liquid, which is then stable with respect to 
the pinwheel dimerization. On the other hand, when $g=0.3$, an energy minimum appears for $\delta t \approx 0.3$, indicating that the pinwheel valence-bond-solid state 
is energetically more convenient than the Dirac spin liquid. The behavior of the energy landscape indicates a second order transition between the spin liquid and the 
pinwheel valence-bond solid. Nevertheless, the lowest-energy variational state found by an unbiased optimization in the $(2,2)_a$ supercell yields the same dimerization 
pattern of the $(2,1)_a$ supercell shown in Fig.~\ref{fig:6x6ene}(b). This fact indicates that the {\it pinwheel} and {\it diamond} patterns with $(2,2)_a$ periodicity, 
proposed in other works as emerging from the monopole condensation of the Dirac spin liquid on the kagome lattice~\cite{song2019,seifert2023}, do not actually describe 
the optimal distortions observed in the present spin-phonon model. 

\section{Conclusion}

In this work, we investigated the potential spin-Peierls instability of the $U(1)$ Dirac spin liquid on the kagome lattice in presence of a coupling between spins and 
lattice distortions. Our analysis is based on a spin-phonon Su-Schrieffer-Heeger Heisenberg Hamiltonian in which the antiferromagnetic exchange interaction between the 
$S=1/2$ spins depends linearly on relative site displacements, the latter being described by a set of uncoupled harmonic oscillators (Einstein phonons). The full quantum 
dynamics of the system is taken into account by means of a variational Monte Carlo approach in which both spins and phonons are treated as quantum mechanical degrees of 
freedom. Even though we have focused our attention on optical phonons, we do not expect a drastically different scenario when considering an acoustic dispersion. In fact, 
since all the distortions found in the numerical simulations possess finite momenta, the difference between optical and acoustic dispersions is not expected to play a 
relevant role. Furthermore, given the substantial energy gain provided by distorted states that cannot be continuously connected to the Dirac spin liquid, it appears 
rather unplausible to obtain a continuous transition between the spin liquid and the valence-bond-solid phases.

Scanning through several supercells and patterns of distortions, we identified the lowest-energy valence-bond ordered states induced by the spin-phonon coupling, which 
display a large density of perfect hexagons~\cite{marston1991}, i.e., (rotationally symmetric) hexagonal plaquettes with strong antiferromagnetic correlations at the 
edges. Other valence-bond ordered states, including those associated to the condensation of monopoles~\cite{song2019,seifert2023}, are found to be energetically less 
convenient. The most important outcome of the present study is the observation that the transition towards valence-bond order takes place at finite values of the spin-phonon 
interaction, which implies that the $U(1)$ Dirac spin liquid is a {\it stable} phase. Furthermore, the analysis of the variational energies, supported by considerations 
about the gauge-invariant fluxes characterizing the variational {\it Ans\"atze}, indicate that the transition betweeen the Dirac spin liquid and the optimal valence-bond 
ordered states is of the first order.

\section*{Acknowledgments}
We thank Y. Iqbal, K. Penc, and J. Schnack for useful discussions. F.F. and R.V. acknowledge support by the Deutsche Forschungsgemeinschaft (DFG, German Research Foundation) 
for funding through TRR 288 -- 422213477 (project A05).


\begin{thebibliography}{57}
\expandafter\ifx\csname natexlab\endcsname\relax\def\natexlab#1{#1}\fi
\expandafter\ifx\csname bibnamefont\endcsname\relax
  \def\bibnamefont#1{#1}\fi
\expandafter\ifx\csname bibfnamefont\endcsname\relax
  \def\bibfnamefont#1{#1}\fi
\expandafter\ifx\csname citenamefont\endcsname\relax
  \def\citenamefont#1{#1}\fi
\expandafter\ifx\csname url\endcsname\relax
  \def\url#1{\texttt{#1}}\fi
\expandafter\ifx\csname urlprefix\endcsname\relax\def\urlprefix{URL }\fi
\providecommand{\bibinfo}[2]{#2}
\providecommand{\eprint}[2][]{\url{#2}}

\bibitem[{\citenamefont{Misguich and Lhuillier}(2005)}]{misguich_review}
\bibinfo{author}{\bibfnamefont{G.}~\bibnamefont{Misguich}} \bibnamefont{and}
  \bibinfo{author}{\bibfnamefont{C.}~\bibnamefont{Lhuillier}},
  \emph{\bibinfo{title}{Two-dimensional Quantum Antiferromagnets}}
  (\bibinfo{publisher}{World Scientific}, \bibinfo{year}{2005}), pp.
  \bibinfo{pages}{229--306}, Frustrated Spin Systems, ISBN
  \bibinfo{isbn}{978-981-256-091-9},
  \urlprefix\url{https://doi.org/10.1142/9789812567819\_0005}.

\bibitem[{\citenamefont{Chalker and Eastmond}(1992)}]{chalker1992}
\bibinfo{author}{\bibfnamefont{J.~T.} \bibnamefont{Chalker}} \bibnamefont{and}
  \bibinfo{author}{\bibfnamefont{J.~F.~G.} \bibnamefont{Eastmond}},
  \bibinfo{journal}{Phys. Rev. B} \textbf{\bibinfo{volume}{46}},
  \bibinfo{pages}{14201} (\bibinfo{year}{1992}),
  \urlprefix\url{https://link.aps.org/doi/10.1103/PhysRevB.46.14201}.

\bibitem[{\citenamefont{Singh and Huse}(1992)}]{singh1992}
\bibinfo{author}{\bibfnamefont{R.~R.~P.} \bibnamefont{Singh}} \bibnamefont{and}
  \bibinfo{author}{\bibfnamefont{D.~A.} \bibnamefont{Huse}},
  \bibinfo{journal}{Phys. Rev. Lett.} \textbf{\bibinfo{volume}{68}},
  \bibinfo{pages}{1766} (\bibinfo{year}{1992}),
  \urlprefix\url{https://link.aps.org/doi/10.1103/PhysRevLett.68.1766}.

\bibitem[{\citenamefont{Leung and Elser}(1993)}]{leung1993}
\bibinfo{author}{\bibfnamefont{P.~W.} \bibnamefont{Leung}} \bibnamefont{and}
  \bibinfo{author}{\bibfnamefont{V.}~\bibnamefont{Elser}},
  \bibinfo{journal}{Phys. Rev. B} \textbf{\bibinfo{volume}{47}},
  \bibinfo{pages}{5459} (\bibinfo{year}{1993}),
  \urlprefix\url{https://link.aps.org/doi/10.1103/PhysRevB.47.5459}.

\bibitem[{\citenamefont{Sachdev}(1992)}]{sachdev1991}
\bibinfo{author}{\bibfnamefont{S.}~\bibnamefont{Sachdev}},
  \bibinfo{journal}{Phys. Rev. B} \textbf{\bibinfo{volume}{45}},
  \bibinfo{pages}{12377} (\bibinfo{year}{1992}),
  \urlprefix\url{https://link.aps.org/doi/10.1103/PhysRevB.45.12377}.

\bibitem[{\citenamefont{Marston and Zeng}(1991)}]{marston1991}
\bibinfo{author}{\bibfnamefont{J.~B.} \bibnamefont{Marston}} \bibnamefont{and}
  \bibinfo{author}{\bibfnamefont{C.}~\bibnamefont{Zeng}},
  \bibinfo{journal}{Journal of Applied Physics} \textbf{\bibinfo{volume}{69}},
  \bibinfo{pages}{5962} (\bibinfo{year}{1991}), ISSN \bibinfo{issn}{0021-8979},
  \eprint{https://pubs.aip.org/aip/jap/article-pdf/69/8/5962/10574767/5962\_1\_online.pdf},
  \urlprefix\url{https://doi.org/10.1063/1.347830}.

\bibitem[{\citenamefont{Nikolic and Senthil}(2003)}]{nikolic2003}
\bibinfo{author}{\bibfnamefont{P.}~\bibnamefont{Nikolic}} \bibnamefont{and}
  \bibinfo{author}{\bibfnamefont{T.}~\bibnamefont{Senthil}},
  \bibinfo{journal}{Phys. Rev. B} \textbf{\bibinfo{volume}{68}},
  \bibinfo{pages}{214415} (\bibinfo{year}{2003}),
  \urlprefix\url{https://link.aps.org/doi/10.1103/PhysRevB.68.214415}.

\bibitem[{\citenamefont{Singh and Huse}(2007)}]{singh2007}
\bibinfo{author}{\bibfnamefont{R.~R.~P.} \bibnamefont{Singh}} \bibnamefont{and}
  \bibinfo{author}{\bibfnamefont{D.~A.} \bibnamefont{Huse}},
  \bibinfo{journal}{Phys. Rev. B} \textbf{\bibinfo{volume}{76}},
  \bibinfo{pages}{180407} (\bibinfo{year}{2007}),
  \urlprefix\url{https://link.aps.org/doi/10.1103/PhysRevB.76.180407}.

\bibitem[{\citenamefont{Poilblanc and Misguich}(2011)}]{poilblanc2011}
\bibinfo{author}{\bibfnamefont{D.}~\bibnamefont{Poilblanc}} \bibnamefont{and}
  \bibinfo{author}{\bibfnamefont{G.}~\bibnamefont{Misguich}},
  \bibinfo{journal}{Phys. Rev. B} \textbf{\bibinfo{volume}{84}},
  \bibinfo{pages}{214401} (\bibinfo{year}{2011}),
  \urlprefix\url{https://link.aps.org/doi/10.1103/PhysRevB.84.214401}.

\bibitem[{\citenamefont{Singh and Huse}(2008)}]{singh2008}
\bibinfo{author}{\bibfnamefont{R.~R.~P.} \bibnamefont{Singh}} \bibnamefont{and}
  \bibinfo{author}{\bibfnamefont{D.~A.} \bibnamefont{Huse}},
  \bibinfo{journal}{Phys. Rev. B} \textbf{\bibinfo{volume}{77}},
  \bibinfo{pages}{144415} (\bibinfo{year}{2008}),
  \urlprefix\url{https://link.aps.org/doi/10.1103/PhysRevB.77.144415}.

\bibitem[{\citenamefont{Poilblanc et~al.}(2010)\citenamefont{Poilblanc,
  Mambrini, and Schwandt}}]{poilblanc2010}
\bibinfo{author}{\bibfnamefont{D.}~\bibnamefont{Poilblanc}},
  \bibinfo{author}{\bibfnamefont{M.}~\bibnamefont{Mambrini}}, \bibnamefont{and}
  \bibinfo{author}{\bibfnamefont{D.}~\bibnamefont{Schwandt}},
  \bibinfo{journal}{Phys. Rev. B} \textbf{\bibinfo{volume}{81}},
  \bibinfo{pages}{180402} (\bibinfo{year}{2010}),
  \urlprefix\url{https://link.aps.org/doi/10.1103/PhysRevB.81.180402}.

\bibitem[{\citenamefont{Lecheminant et~al.}(1997)\citenamefont{Lecheminant,
  Bernu, Lhuillier, Pierre, and Sindzingre}}]{lecheminant1997}
\bibinfo{author}{\bibfnamefont{P.}~\bibnamefont{Lecheminant}},
  \bibinfo{author}{\bibfnamefont{B.}~\bibnamefont{Bernu}},
  \bibinfo{author}{\bibfnamefont{C.}~\bibnamefont{Lhuillier}},
  \bibinfo{author}{\bibfnamefont{L.}~\bibnamefont{Pierre}}, \bibnamefont{and}
  \bibinfo{author}{\bibfnamefont{P.}~\bibnamefont{Sindzingre}},
  \bibinfo{journal}{Phys. Rev. B} \textbf{\bibinfo{volume}{56}},
  \bibinfo{pages}{2521} (\bibinfo{year}{1997}),
  \urlprefix\url{https://link.aps.org/doi/10.1103/PhysRevB.56.2521}.

\bibitem[{\citenamefont{Mila}(1998)}]{mila1998}
\bibinfo{author}{\bibfnamefont{F.}~\bibnamefont{Mila}}, \bibinfo{journal}{Phys.
  Rev. Lett.} \textbf{\bibinfo{volume}{81}}, \bibinfo{pages}{2356}
  (\bibinfo{year}{1998}),
  \urlprefix\url{https://link.aps.org/doi/10.1103/PhysRevLett.81.2356}.

\bibitem[{\citenamefont{Mambrini and Mila}(2000)}]{mambrini2000}
\bibinfo{author}{\bibfnamefont{M.}~\bibnamefont{Mambrini}} \bibnamefont{and}
  \bibinfo{author}{\bibfnamefont{F.}~\bibnamefont{Mila}}, \bibinfo{journal}{The
  European Physical Journal B - Condensed Matter and Complex Systems}
  \textbf{\bibinfo{volume}{17}}, \bibinfo{pages}{651} (\bibinfo{year}{2000}),
  \urlprefix\url{https://doi.org/10.1007/PL00011071}.

\bibitem[{\citenamefont{L\"auchli et~al.}(2019)\citenamefont{L\"auchli, Sudan,
  and Moessner}}]{laeuchli2019}
\bibinfo{author}{\bibfnamefont{A.~M.} \bibnamefont{L\"auchli}},
  \bibinfo{author}{\bibfnamefont{J.}~\bibnamefont{Sudan}}, \bibnamefont{and}
  \bibinfo{author}{\bibfnamefont{R.}~\bibnamefont{Moessner}},
  \bibinfo{journal}{Phys. Rev. B} \textbf{\bibinfo{volume}{100}},
  \bibinfo{pages}{155142} (\bibinfo{year}{2019}),
  \urlprefix\url{https://link.aps.org/doi/10.1103/PhysRevB.100.155142}.

\bibitem[{\citenamefont{Zeng and Elser}(1995)}]{zeng1995}
\bibinfo{author}{\bibfnamefont{C.}~\bibnamefont{Zeng}} \bibnamefont{and}
  \bibinfo{author}{\bibfnamefont{V.}~\bibnamefont{Elser}},
  \bibinfo{journal}{Phys. Rev. B} \textbf{\bibinfo{volume}{51}},
  \bibinfo{pages}{8318} (\bibinfo{year}{1995}),
  \urlprefix\url{https://link.aps.org/doi/10.1103/PhysRevB.51.8318}.

\bibitem[{\citenamefont{Jiang et~al.}(2008)\citenamefont{Jiang, Weng, and
  Sheng}}]{jiang2008}
\bibinfo{author}{\bibfnamefont{H.~C.} \bibnamefont{Jiang}},
  \bibinfo{author}{\bibfnamefont{Z.~Y.} \bibnamefont{Weng}}, \bibnamefont{and}
  \bibinfo{author}{\bibfnamefont{D.~N.} \bibnamefont{Sheng}},
  \bibinfo{journal}{Phys. Rev. Lett.} \textbf{\bibinfo{volume}{101}},
  \bibinfo{pages}{117203} (\bibinfo{year}{2008}),
  \urlprefix\url{https://link.aps.org/doi/10.1103/PhysRevLett.101.117203}.

\bibitem[{\citenamefont{Yan et~al.}(2011)\citenamefont{Yan, Huse, and
  White}}]{yan2011}
\bibinfo{author}{\bibfnamefont{S.}~\bibnamefont{Yan}},
  \bibinfo{author}{\bibfnamefont{D.~A.} \bibnamefont{Huse}}, \bibnamefont{and}
  \bibinfo{author}{\bibfnamefont{S.~R.} \bibnamefont{White}},
  \bibinfo{journal}{Science} \textbf{\bibinfo{volume}{332}},
  \bibinfo{pages}{1173} (\bibinfo{year}{2011}),
  \urlprefix\url{https://www.science.org/doi/abs/10.1126/science.1201080}.

\bibitem[{\citenamefont{Depenbrock et~al.}(2012)\citenamefont{Depenbrock,
  McCulloch, and Schollw\"ock}}]{depenbrock2012}
\bibinfo{author}{\bibfnamefont{S.}~\bibnamefont{Depenbrock}},
  \bibinfo{author}{\bibfnamefont{I.~P.} \bibnamefont{McCulloch}},
  \bibnamefont{and}
  \bibinfo{author}{\bibfnamefont{U.}~\bibnamefont{Schollw\"ock}},
  \bibinfo{journal}{Phys. Rev. Lett.} \textbf{\bibinfo{volume}{109}},
  \bibinfo{pages}{067201} (\bibinfo{year}{2012}),
  \urlprefix\url{https://link.aps.org/doi/10.1103/PhysRevLett.109.067201}.

\bibitem[{\citenamefont{Iqbal et~al.}(2013)\citenamefont{Iqbal, Becca, Sorella,
  and Poilblanc}}]{iqbal2013}
\bibinfo{author}{\bibfnamefont{Y.}~\bibnamefont{Iqbal}},
  \bibinfo{author}{\bibfnamefont{F.}~\bibnamefont{Becca}},
  \bibinfo{author}{\bibfnamefont{S.}~\bibnamefont{Sorella}}, \bibnamefont{and}
  \bibinfo{author}{\bibfnamefont{D.}~\bibnamefont{Poilblanc}},
  \bibinfo{journal}{Phys. Rev. B} \textbf{\bibinfo{volume}{87}},
  \bibinfo{pages}{060405} (\bibinfo{year}{2013}),
  \urlprefix\url{https://link.aps.org/doi/10.1103/PhysRevB.87.060405}.

\bibitem[{\citenamefont{Ran et~al.}(2007)\citenamefont{Ran, Hermele, Lee, and
  Wen}}]{ran2007}
\bibinfo{author}{\bibfnamefont{Y.}~\bibnamefont{Ran}},
  \bibinfo{author}{\bibfnamefont{M.}~\bibnamefont{Hermele}},
  \bibinfo{author}{\bibfnamefont{P.~A.} \bibnamefont{Lee}}, \bibnamefont{and}
  \bibinfo{author}{\bibfnamefont{X.-G.} \bibnamefont{Wen}},
  \bibinfo{journal}{Phys. Rev. Lett.} \textbf{\bibinfo{volume}{98}},
  \bibinfo{pages}{117205} (\bibinfo{year}{2007}),
  \urlprefix\url{https://link.aps.org/doi/10.1103/PhysRevLett.98.117205}.

\bibitem[{\citenamefont{Wen}(2002)}]{wen2002}
\bibinfo{author}{\bibfnamefont{X.-G.} \bibnamefont{Wen}},
  \bibinfo{journal}{Phys. Rev. B} \textbf{\bibinfo{volume}{65}},
  \bibinfo{pages}{165113} (\bibinfo{year}{2002}),
  \urlprefix\url{https://link.aps.org/doi/10.1103/PhysRevB.65.165113}.

\bibitem[{\citenamefont{Iqbal et~al.}(2012)\citenamefont{Iqbal, Becca, and
  Poilblanc}}]{iqbal2012}
\bibinfo{author}{\bibfnamefont{Y.}~\bibnamefont{Iqbal}},
  \bibinfo{author}{\bibfnamefont{F.}~\bibnamefont{Becca}}, \bibnamefont{and}
  \bibinfo{author}{\bibfnamefont{D.}~\bibnamefont{Poilblanc}},
  \bibinfo{journal}{New Journal of Physics} \textbf{\bibinfo{volume}{14}},
  \bibinfo{pages}{115031} (\bibinfo{year}{2012}),
  \urlprefix\url{https://dx.doi.org/10.1088/1367-2630/14/11/115031}.

\bibitem[{\citenamefont{Iqbal et~al.}(2011)\citenamefont{Iqbal, Becca, and
  Poilblanc}}]{iqbal2011z2}
\bibinfo{author}{\bibfnamefont{Y.}~\bibnamefont{Iqbal}},
  \bibinfo{author}{\bibfnamefont{F.}~\bibnamefont{Becca}}, \bibnamefont{and}
  \bibinfo{author}{\bibfnamefont{D.}~\bibnamefont{Poilblanc}},
  \bibinfo{journal}{Phys. Rev. B} \textbf{\bibinfo{volume}{84}},
  \bibinfo{pages}{020407} (\bibinfo{year}{2011}),
  \urlprefix\url{https://link.aps.org/doi/10.1103/PhysRevB.84.020407}.

\bibitem[{\citenamefont{He et~al.}(2017)\citenamefont{He, Zaletel, Oshikawa,
  and Pollmann}}]{he2017}
\bibinfo{author}{\bibfnamefont{Y.-C.} \bibnamefont{He}},
  \bibinfo{author}{\bibfnamefont{M.~P.} \bibnamefont{Zaletel}},
  \bibinfo{author}{\bibfnamefont{M.}~\bibnamefont{Oshikawa}}, \bibnamefont{and}
  \bibinfo{author}{\bibfnamefont{F.}~\bibnamefont{Pollmann}},
  \bibinfo{journal}{Phys. Rev. X} \textbf{\bibinfo{volume}{7}},
  \bibinfo{pages}{031020} (\bibinfo{year}{2017}),
  \urlprefix\url{https://link.aps.org/doi/10.1103/PhysRevX.7.031020}.

\bibitem[{\citenamefont{Zhu et~al.}(2018)\citenamefont{Zhu, Chen, He, and
  Witczak-Krempa}}]{zhu2018}
\bibinfo{author}{\bibfnamefont{W.}~\bibnamefont{Zhu}},
  \bibinfo{author}{\bibfnamefont{X.}~\bibnamefont{Chen}},
  \bibinfo{author}{\bibfnamefont{Y.-C.} \bibnamefont{He}}, \bibnamefont{and}
  \bibinfo{author}{\bibfnamefont{W.}~\bibnamefont{Witczak-Krempa}},
  \bibinfo{journal}{Science Advances} \textbf{\bibinfo{volume}{4}},
  \bibinfo{pages}{eaat5535} (\bibinfo{year}{2018}),
  \urlprefix\url{https://www.science.org/doi/abs/10.1126/sciadv.aat5535}.

\bibitem[{\citenamefont{Ferrari
  et~al.}(2021{\natexlab{a}})\citenamefont{Ferrari, Parola, and
  Becca}}]{ferrari2021cylinders}
\bibinfo{author}{\bibfnamefont{F.}~\bibnamefont{Ferrari}},
  \bibinfo{author}{\bibfnamefont{A.}~\bibnamefont{Parola}}, \bibnamefont{and}
  \bibinfo{author}{\bibfnamefont{F.}~\bibnamefont{Becca}},
  \bibinfo{journal}{Phys. Rev. B} \textbf{\bibinfo{volume}{103}},
  \bibinfo{pages}{195140} (\bibinfo{year}{2021}{\natexlab{a}}),
  \urlprefix\url{https://link.aps.org/doi/10.1103/PhysRevB.103.195140}.

\bibitem[{\citenamefont{Liao et~al.}(2017)\citenamefont{Liao, Xie, Chen, Liu,
  Xie, Huang, Normand, and Xiang}}]{liao2017}
\bibinfo{author}{\bibfnamefont{H.~J.} \bibnamefont{Liao}},
  \bibinfo{author}{\bibfnamefont{Z.~Y.} \bibnamefont{Xie}},
  \bibinfo{author}{\bibfnamefont{J.}~\bibnamefont{Chen}},
  \bibinfo{author}{\bibfnamefont{Z.~Y.} \bibnamefont{Liu}},
  \bibinfo{author}{\bibfnamefont{H.~D.} \bibnamefont{Xie}},
  \bibinfo{author}{\bibfnamefont{R.~Z.} \bibnamefont{Huang}},
  \bibinfo{author}{\bibfnamefont{B.}~\bibnamefont{Normand}}, \bibnamefont{and}
  \bibinfo{author}{\bibfnamefont{T.}~\bibnamefont{Xiang}},
  \bibinfo{journal}{Phys. Rev. Lett.} \textbf{\bibinfo{volume}{118}},
  \bibinfo{pages}{137202} (\bibinfo{year}{2017}),
  \urlprefix\url{https://link.aps.org/doi/10.1103/PhysRevLett.118.137202}.

\bibitem[{\citenamefont{Song et~al.}(2019)\citenamefont{Song, Wang, Vishwanath,
  and He}}]{song2019}
\bibinfo{author}{\bibfnamefont{X.-Y.} \bibnamefont{Song}},
  \bibinfo{author}{\bibfnamefont{C.}~\bibnamefont{Wang}},
  \bibinfo{author}{\bibfnamefont{A.}~\bibnamefont{Vishwanath}},
  \bibnamefont{and} \bibinfo{author}{\bibfnamefont{Y.-C.} \bibnamefont{He}},
  \bibinfo{journal}{Nature Communications} \textbf{\bibinfo{volume}{10}},
  \bibinfo{pages}{4254} (\bibinfo{year}{2019}),
  \urlprefix\url{https://doi.org/10.1038/s41467-019-11727-3}.

\bibitem[{\citenamefont{Hastings}(2000)}]{hastings2000}
\bibinfo{author}{\bibfnamefont{M.~B.} \bibnamefont{Hastings}},
  \bibinfo{journal}{Phys. Rev. B} \textbf{\bibinfo{volume}{63}},
  \bibinfo{pages}{014413} (\bibinfo{year}{2000}),
  \urlprefix\url{https://link.aps.org/doi/10.1103/PhysRevB.63.014413}.

\bibitem[{\citenamefont{Hermele et~al.}(2008)\citenamefont{Hermele, Ran, Lee,
  and Wen}}]{hermele2008}
\bibinfo{author}{\bibfnamefont{M.}~\bibnamefont{Hermele}},
  \bibinfo{author}{\bibfnamefont{Y.}~\bibnamefont{Ran}},
  \bibinfo{author}{\bibfnamefont{P.~A.} \bibnamefont{Lee}}, \bibnamefont{and}
  \bibinfo{author}{\bibfnamefont{X.-G.} \bibnamefont{Wen}},
  \bibinfo{journal}{Phys. Rev. B} \textbf{\bibinfo{volume}{77}},
  \bibinfo{pages}{224413} (\bibinfo{year}{2008}),
  \urlprefix\url{https://link.aps.org/doi/10.1103/PhysRevB.77.224413}.

\bibitem[{\citenamefont{Song et~al.}(2020)\citenamefont{Song, He, Vishwanath,
  and Wang}}]{song2020}
\bibinfo{author}{\bibfnamefont{X.-Y.} \bibnamefont{Song}},
  \bibinfo{author}{\bibfnamefont{Y.-C.} \bibnamefont{He}},
  \bibinfo{author}{\bibfnamefont{A.}~\bibnamefont{Vishwanath}},
  \bibnamefont{and} \bibinfo{author}{\bibfnamefont{C.}~\bibnamefont{Wang}},
  \bibinfo{journal}{Phys. Rev. X} \textbf{\bibinfo{volume}{10}},
  \bibinfo{pages}{011033} (\bibinfo{year}{2020}),
  \urlprefix\url{https://link.aps.org/doi/10.1103/PhysRevX.10.011033}.

\bibitem[{\citenamefont{Gong et~al.}(2015)\citenamefont{Gong, Zhu, Balents, and
  Sheng}}]{gong2015}
\bibinfo{author}{\bibfnamefont{S.-S.} \bibnamefont{Gong}},
  \bibinfo{author}{\bibfnamefont{W.}~\bibnamefont{Zhu}},
  \bibinfo{author}{\bibfnamefont{L.}~\bibnamefont{Balents}}, \bibnamefont{and}
  \bibinfo{author}{\bibfnamefont{D.~N.} \bibnamefont{Sheng}},
  \bibinfo{journal}{Phys. Rev. B} \textbf{\bibinfo{volume}{91}},
  \bibinfo{pages}{075112} (\bibinfo{year}{2015}),
  \urlprefix\url{https://link.aps.org/doi/10.1103/PhysRevB.91.075112}.

\bibitem[{\citenamefont{Kolley et~al.}(2015)\citenamefont{Kolley, Depenbrock,
  McCulloch, Schollw\"ock, and Alba}}]{kolley2015}
\bibinfo{author}{\bibfnamefont{F.}~\bibnamefont{Kolley}},
  \bibinfo{author}{\bibfnamefont{S.}~\bibnamefont{Depenbrock}},
  \bibinfo{author}{\bibfnamefont{I.~P.} \bibnamefont{McCulloch}},
  \bibinfo{author}{\bibfnamefont{U.}~\bibnamefont{Schollw\"ock}},
  \bibnamefont{and} \bibinfo{author}{\bibfnamefont{V.}~\bibnamefont{Alba}},
  \bibinfo{journal}{Phys. Rev. B} \textbf{\bibinfo{volume}{91}},
  \bibinfo{pages}{104418} (\bibinfo{year}{2015}),
  \urlprefix\url{https://link.aps.org/doi/10.1103/PhysRevB.91.104418}.

\bibitem[{\citenamefont{Iqbal et~al.}(2021)\citenamefont{Iqbal, Ferrari,
  Chauhan, Parola, Poilblanc, and Becca}}]{iqbal2021}
\bibinfo{author}{\bibfnamefont{Y.}~\bibnamefont{Iqbal}},
  \bibinfo{author}{\bibfnamefont{F.}~\bibnamefont{Ferrari}},
  \bibinfo{author}{\bibfnamefont{A.}~\bibnamefont{Chauhan}},
  \bibinfo{author}{\bibfnamefont{A.}~\bibnamefont{Parola}},
  \bibinfo{author}{\bibfnamefont{D.}~\bibnamefont{Poilblanc}},
  \bibnamefont{and} \bibinfo{author}{\bibfnamefont{F.}~\bibnamefont{Becca}},
  \bibinfo{journal}{Phys. Rev. B} \textbf{\bibinfo{volume}{104}},
  \bibinfo{pages}{144406} (\bibinfo{year}{2021}),
  \urlprefix\url{https://link.aps.org/doi/10.1103/PhysRevB.104.144406}.

\bibitem[{\citenamefont{Wietek and L\"auchli}(2020)}]{wietek2020}
\bibinfo{author}{\bibfnamefont{A.}~\bibnamefont{Wietek}} \bibnamefont{and}
  \bibinfo{author}{\bibfnamefont{A.~M.} \bibnamefont{L\"auchli}},
  \bibinfo{journal}{Phys. Rev. B} \textbf{\bibinfo{volume}{102}},
  \bibinfo{pages}{020411} (\bibinfo{year}{2020}),
  \urlprefix\url{https://link.aps.org/doi/10.1103/PhysRevB.102.020411}.

\bibitem[{\citenamefont{Kiese et~al.}(2023)\citenamefont{Kiese, Ferrari,
  Astrakhantsev, Niggemann, Ghosh, M\"uller, Thomale, Neupert, Reuther, Gingras
  et~al.}}]{kiese2023}
\bibinfo{author}{\bibfnamefont{D.}~\bibnamefont{Kiese}},
  \bibinfo{author}{\bibfnamefont{F.}~\bibnamefont{Ferrari}},
  \bibinfo{author}{\bibfnamefont{N.}~\bibnamefont{Astrakhantsev}},
  \bibinfo{author}{\bibfnamefont{N.}~\bibnamefont{Niggemann}},
  \bibinfo{author}{\bibfnamefont{P.}~\bibnamefont{Ghosh}},
  \bibinfo{author}{\bibfnamefont{T.}~\bibnamefont{M\"uller}},
  \bibinfo{author}{\bibfnamefont{R.}~\bibnamefont{Thomale}},
  \bibinfo{author}{\bibfnamefont{T.}~\bibnamefont{Neupert}},
  \bibinfo{author}{\bibfnamefont{J.}~\bibnamefont{Reuther}},
  \bibinfo{author}{\bibfnamefont{M.~J.~P.} \bibnamefont{Gingras}},
  \bibnamefont{et~al.}, \bibinfo{journal}{Phys. Rev. Res.}
  \textbf{\bibinfo{volume}{5}}, \bibinfo{pages}{L012025}
  (\bibinfo{year}{2023}),
  \urlprefix\url{https://link.aps.org/doi/10.1103/PhysRevResearch.5.L012025}.

\bibitem[{\citenamefont{Cross and Fisher}(1979)}]{cross1979}
\bibinfo{author}{\bibfnamefont{M.~C.} \bibnamefont{Cross}} \bibnamefont{and}
  \bibinfo{author}{\bibfnamefont{D.~S.} \bibnamefont{Fisher}},
  \bibinfo{journal}{Phys. Rev. B} \textbf{\bibinfo{volume}{19}},
  \bibinfo{pages}{402} (\bibinfo{year}{1979}),
  \urlprefix\url{https://link.aps.org/doi/10.1103/PhysRevB.19.402}.

\bibitem[{\citenamefont{Pytte}(1974)}]{pytte1974}
\bibinfo{author}{\bibfnamefont{E.}~\bibnamefont{Pytte}},
  \bibinfo{journal}{Phys. Rev. B} \textbf{\bibinfo{volume}{10}},
  \bibinfo{pages}{4637} (\bibinfo{year}{1974}),
  \urlprefix\url{https://link.aps.org/doi/10.1103/PhysRevB.10.4637}.

\bibitem[{\citenamefont{Uhrig}(1998)}]{uhrig1998}
\bibinfo{author}{\bibfnamefont{G.~S.} \bibnamefont{Uhrig}},
  \bibinfo{journal}{Phys. Rev. B} \textbf{\bibinfo{volume}{57}},
  \bibinfo{pages}{R14004} (\bibinfo{year}{1998}),
  \urlprefix\url{https://link.aps.org/doi/10.1103/PhysRevB.57.R14004}.

\bibitem[{\citenamefont{Bursill et~al.}(1999)\citenamefont{Bursill, McKenzie,
  and Hamer}}]{bursill1999}
\bibinfo{author}{\bibfnamefont{R.~J.} \bibnamefont{Bursill}},
  \bibinfo{author}{\bibfnamefont{R.~H.} \bibnamefont{McKenzie}},
  \bibnamefont{and} \bibinfo{author}{\bibfnamefont{C.~J.} \bibnamefont{Hamer}},
  \bibinfo{journal}{Phys. Rev. Lett.} \textbf{\bibinfo{volume}{83}},
  \bibinfo{pages}{408} (\bibinfo{year}{1999}),
  \urlprefix\url{https://link.aps.org/doi/10.1103/PhysRevLett.83.408}.

\bibitem[{\citenamefont{Sandvik and Campbell}(1999)}]{sandvik1999}
\bibinfo{author}{\bibfnamefont{A.~W.} \bibnamefont{Sandvik}} \bibnamefont{and}
  \bibinfo{author}{\bibfnamefont{D.~K.} \bibnamefont{Campbell}},
  \bibinfo{journal}{Phys. Rev. Lett.} \textbf{\bibinfo{volume}{83}},
  \bibinfo{pages}{195} (\bibinfo{year}{1999}),
  \urlprefix\url{https://link.aps.org/doi/10.1103/PhysRevLett.83.195}.

\bibitem[{\citenamefont{Ferrari et~al.}(2020)\citenamefont{Ferrari,
  Valent\'{\i}, and Becca}}]{ferrari2020}
\bibinfo{author}{\bibfnamefont{F.}~\bibnamefont{Ferrari}},
  \bibinfo{author}{\bibfnamefont{R.}~\bibnamefont{Valent\'{\i}}},
  \bibnamefont{and} \bibinfo{author}{\bibfnamefont{F.}~\bibnamefont{Becca}},
  \bibinfo{journal}{Phys. Rev. B} \textbf{\bibinfo{volume}{102}},
  \bibinfo{pages}{125149} (\bibinfo{year}{2020}),
  \urlprefix\url{https://link.aps.org/doi/10.1103/PhysRevB.102.125149}.

\bibitem[{\citenamefont{Ye et~al.}(2020)\citenamefont{Ye, Fernandes, and
  Perkins}}]{ye2020}
\bibinfo{author}{\bibfnamefont{M.}~\bibnamefont{Ye}},
  \bibinfo{author}{\bibfnamefont{R.~M.} \bibnamefont{Fernandes}},
  \bibnamefont{and} \bibinfo{author}{\bibfnamefont{N.~B.}
  \bibnamefont{Perkins}}, \bibinfo{journal}{Phys. Rev. Res.}
  \textbf{\bibinfo{volume}{2}}, \bibinfo{pages}{033180} (\bibinfo{year}{2020}),
  \urlprefix\url{https://link.aps.org/doi/10.1103/PhysRevResearch.2.033180}.

\bibitem[{\citenamefont{Metavitsiadis et~al.}(2022)\citenamefont{Metavitsiadis,
  Natori, Knolle, and Brenig}}]{metavitsiadis2022}
\bibinfo{author}{\bibfnamefont{A.}~\bibnamefont{Metavitsiadis}},
  \bibinfo{author}{\bibfnamefont{W.}~\bibnamefont{Natori}},
  \bibinfo{author}{\bibfnamefont{J.}~\bibnamefont{Knolle}}, \bibnamefont{and}
  \bibinfo{author}{\bibfnamefont{W.}~\bibnamefont{Brenig}},
  \bibinfo{journal}{Phys. Rev. B} \textbf{\bibinfo{volume}{105}},
  \bibinfo{pages}{165151} (\bibinfo{year}{2022}),
  \urlprefix\url{https://link.aps.org/doi/10.1103/PhysRevB.105.165151}.

\bibitem[{\citenamefont{Singh et~al.}(2023)\citenamefont{Singh, Stavropoulos,
  and Perkins}}]{singh2023}
\bibinfo{author}{\bibfnamefont{S.}~\bibnamefont{Singh}},
  \bibinfo{author}{\bibfnamefont{P.~P.} \bibnamefont{Stavropoulos}},
  \bibnamefont{and} \bibinfo{author}{\bibfnamefont{N.~B.}
  \bibnamefont{Perkins}}, \bibinfo{journal}{Phys. Rev. B}
  \textbf{\bibinfo{volume}{107}}, \bibinfo{pages}{214428}
  (\bibinfo{year}{2023}),
  \urlprefix\url{https://link.aps.org/doi/10.1103/PhysRevB.107.214428}.

\bibitem[{\citenamefont{Ferrari
  et~al.}(2021{\natexlab{b}})\citenamefont{Ferrari, Valent\'{\i}, and
  Becca}}]{ferrari2021}
\bibinfo{author}{\bibfnamefont{F.}~\bibnamefont{Ferrari}},
  \bibinfo{author}{\bibfnamefont{R.}~\bibnamefont{Valent\'{\i}}},
  \bibnamefont{and} \bibinfo{author}{\bibfnamefont{F.}~\bibnamefont{Becca}},
  \bibinfo{journal}{Phys. Rev. B} \textbf{\bibinfo{volume}{104}},
  \bibinfo{pages}{035126} (\bibinfo{year}{2021}{\natexlab{b}}),
  \urlprefix\url{https://link.aps.org/doi/10.1103/PhysRevB.104.035126}.

\bibitem[{\citenamefont{Richter et~al.}(2004)\citenamefont{Richter, Derzhko,
  and Schulenburg}}]{richter2004}
\bibinfo{author}{\bibfnamefont{J.}~\bibnamefont{Richter}},
  \bibinfo{author}{\bibfnamefont{O.}~\bibnamefont{Derzhko}}, \bibnamefont{and}
  \bibinfo{author}{\bibfnamefont{J.}~\bibnamefont{Schulenburg}},
  \bibinfo{journal}{Phys. Rev. Lett.} \textbf{\bibinfo{volume}{93}},
  \bibinfo{pages}{107206} (\bibinfo{year}{2004}),
  \urlprefix\url{https://link.aps.org/doi/10.1103/PhysRevLett.93.107206}.

\bibitem[{\citenamefont{Seifert et~al.}(2023)\citenamefont{Seifert, Willsher,
  Drescher, Pollmann, and Knolle}}]{seifert2023}
\bibinfo{author}{\bibfnamefont{U.~F.~P.} \bibnamefont{Seifert}},
  \bibinfo{author}{\bibfnamefont{J.}~\bibnamefont{Willsher}},
  \bibinfo{author}{\bibfnamefont{M.}~\bibnamefont{Drescher}},
  \bibinfo{author}{\bibfnamefont{F.}~\bibnamefont{Pollmann}}, \bibnamefont{and}
  \bibinfo{author}{\bibfnamefont{J.}~\bibnamefont{Knolle}},
  \emph{\bibinfo{title}{Spin-peierls instability of the u(1) dirac spin
  liquid}} (\bibinfo{year}{2023}), \eprint{2307.12295}.

\bibitem[{\citenamefont{Su et~al.}(1979)\citenamefont{Su, Schrieffer, and
  Heeger}}]{su1979}
\bibinfo{author}{\bibfnamefont{W.~P.} \bibnamefont{Su}},
  \bibinfo{author}{\bibfnamefont{J.~R.} \bibnamefont{Schrieffer}},
  \bibnamefont{and} \bibinfo{author}{\bibfnamefont{A.~J.}
  \bibnamefont{Heeger}}, \bibinfo{journal}{Phys. Rev. Lett.}
  \textbf{\bibinfo{volume}{42}}, \bibinfo{pages}{1698} (\bibinfo{year}{1979}),
  \urlprefix\url{https://link.aps.org/doi/10.1103/PhysRevLett.42.1698}.

\bibitem[{\citenamefont{Ferrari et~al.}(2022)\citenamefont{Ferrari, Becca, and
  Valent\'{\i}}}]{ferrari2022}
\bibinfo{author}{\bibfnamefont{F.}~\bibnamefont{Ferrari}},
  \bibinfo{author}{\bibfnamefont{F.}~\bibnamefont{Becca}}, \bibnamefont{and}
  \bibinfo{author}{\bibfnamefont{R.}~\bibnamefont{Valent\'{\i}}},
  \bibinfo{journal}{Phys. Rev. B} \textbf{\bibinfo{volume}{106}},
  \bibinfo{pages}{L081107} (\bibinfo{year}{2022}),
  \urlprefix\url{https://link.aps.org/doi/10.1103/PhysRevB.106.L081107}.

\bibitem[{\citenamefont{Abrikosov}(1965)}]{abrikosov1965}
\bibinfo{author}{\bibfnamefont{A.~A.} \bibnamefont{Abrikosov}},
  \bibinfo{journal}{Physics Physique Fizika} \textbf{\bibinfo{volume}{2}},
  \bibinfo{pages}{5} (\bibinfo{year}{1965}),
  \urlprefix\url{https://link.aps.org/doi/10.1103/PhysicsPhysiqueFizika.2.5}.

\bibitem[{\citenamefont{Savary and Balents}(2016)}]{savary2016}
\bibinfo{author}{\bibfnamefont{L.}~\bibnamefont{Savary}} \bibnamefont{and}
  \bibinfo{author}{\bibfnamefont{L.}~\bibnamefont{Balents}},
  \bibinfo{journal}{Reports on Progress in Physics}
  \textbf{\bibinfo{volume}{80}}, \bibinfo{pages}{016502}
  (\bibinfo{year}{2016}),
  \urlprefix\url{https://doi.org/10.1088/0034-4885/80/1/016502}.

\bibitem[{\citenamefont{Sorella}(2005)}]{sorella2005}
\bibinfo{author}{\bibfnamefont{S.}~\bibnamefont{Sorella}},
  \bibinfo{journal}{Phys. Rev. B} \textbf{\bibinfo{volume}{71}},
  \bibinfo{pages}{241103} (\bibinfo{year}{2005}),
  \urlprefix\url{https://link.aps.org/doi/10.1103/PhysRevB.71.241103}.

\bibitem[{\citenamefont{Lu et~al.}(2011)\citenamefont{Lu, Ran, and
  Lee}}]{lu2011}
\bibinfo{author}{\bibfnamefont{Y.-M.} \bibnamefont{Lu}},
  \bibinfo{author}{\bibfnamefont{Y.}~\bibnamefont{Ran}}, \bibnamefont{and}
  \bibinfo{author}{\bibfnamefont{P.~A.} \bibnamefont{Lee}},
  \bibinfo{journal}{Phys. Rev. B} \textbf{\bibinfo{volume}{83}},
  \bibinfo{pages}{224413} (\bibinfo{year}{2011}),
  \urlprefix\url{https://link.aps.org/doi/10.1103/PhysRevB.83.224413}.

\bibitem[{\citenamefont{Evenbly and Vidal}(2010)}]{evenbly2010}
\bibinfo{author}{\bibfnamefont{G.}~\bibnamefont{Evenbly}} \bibnamefont{and}
  \bibinfo{author}{\bibfnamefont{G.}~\bibnamefont{Vidal}},
  \bibinfo{journal}{Phys. Rev. Lett.} \textbf{\bibinfo{volume}{104}},
  \bibinfo{pages}{187203} (\bibinfo{year}{2010}),
  \urlprefix\url{https://link.aps.org/doi/10.1103/PhysRevLett.104.187203}.

\bibitem[{\citenamefont{Huh et~al.}(2011)\citenamefont{Huh, Punk, and
  Sachdev}}]{huh2011}
\bibinfo{author}{\bibfnamefont{Y.}~\bibnamefont{Huh}},
  \bibinfo{author}{\bibfnamefont{M.}~\bibnamefont{Punk}}, \bibnamefont{and}
  \bibinfo{author}{\bibfnamefont{S.}~\bibnamefont{Sachdev}},
  \bibinfo{journal}{Phys. Rev. B} \textbf{\bibinfo{volume}{84}},
  \bibinfo{pages}{094419} (\bibinfo{year}{2011}),
  \urlprefix\url{https://link.aps.org/doi/10.1103/PhysRevB.84.094419}.

\end{thebibliography}
\end{document}